\documentclass[pra,aps,twocolumn,superscriptaddress,floatfix,nofootinbib,showpacs,longbibliography]{revtex4-1}

\usepackage[utf8]{inputenc}  
\usepackage[T1]{fontenc}     
\usepackage[british]{babel}  
\usepackage[sc,osf]{mathpazo}
\usepackage[scaled=0.86]{berasans}  
\usepackage[colorlinks=true, citecolor=blue, urlcolor=blue]{hyperref}  
\usepackage{scrextend}
\usepackage{graphicx} 
\usepackage[babel]{microtype}  
\usepackage{amsmath,amssymb,amsthm,bm,amsfonts,mathrsfs,bbm} 

\usepackage{xspace}  
\usepackage{pgf,tikz}
\usepackage{xcolor,colortbl}
\usepackage{array}
\usepackage{bigstrut}
\usepackage{mathtools}
\usepackage{braket}
\usepackage{color}
\usepackage{natbib}
\usepackage{slashbox}

\newcommand{\be}{\begin{equation}}
\newcommand{\ee}{\end{equation}}
\newcommand{\ba}{\begin{eqnarray}}
\newcommand{\ea}{\end{eqnarray}}

\begin{document}

\title{Uncertainty principle as post-quantum nonlocality witness for continuous variable multi-mode scenario}

\author{Prathik Cherian J}
\affiliation{Optics and Quantum Information Group, The Institute of Mathematical Sciences, C. I. T. Campus, Taramani, Chennai 600113, India}
\affiliation{Homi Bhabha National Institute, Training School Complex, Anushakti Nagar, Mumbai 400094, India}
\author{Amit Mukherjee}
\affiliation{Optics and Quantum Information Group, The Institute of Mathematical Sciences, C. I. T. Campus, Taramani, Chennai 600113, India}
\affiliation{Homi Bhabha National Institute, Training School Complex, Anushakti Nagar, Mumbai 400094, India}
\author{Arup Roy}
\affiliation{Physics and Applied Mathematics Unit, Indian Statistical Institute, 203 B.T. Road, Kolkata-700108, India}

\author{Some Sankar Bhattacharya}
\affiliation{Physics and Applied Mathematics Unit, Indian Statistical Institute, 203 B.T. Road, Kolkata-700108, India}

\author{Manik Banik}
\affiliation{S.N. Bose National Center for Basic Sciences, Block JD, Sector III, Salt Lake, Kolkata-700098, India}

\begin{abstract}
Uncertainty principle is one of the central concepts in quantum theory. Different forms of this particular principle have been discoursed in various foundational and information theoretic topics. In the discrete input-output scenario the limited nonlocal behavior of quantum theory has been explained by \emph{fine grained} uncertainty relation. On the other hand, in continuous variable paradigm Robertson-Schr\"{o}dinger (RS) uncertainty relation has been used to detect multi-mode entanglement. Here we show that RS uncertainty relation plays an important role to discriminate between quantum and post-quantum nonlocal correlations in multi-mode continuous outcome scenario. We provide a class of $m$-mode post-quantum nonlocal correlations with continuous outcome spectrum.  While nonlocality of the introduced class of correlations is established through Calvalcanti-Foster-Reid-Drummond (CFRD) class of Bell inequalities, RS uncertainty relation detects their post-quantum nature. Our result is a hint towards a wider role of uncertainty principle in the study of nonlocality in continuous variable multi-mode systems. 
\end{abstract}



\maketitle

\section{Introduction}
Nonlocality is one of the most bizarre features of multipartite quantum systems established for the very first time in the seminal paper by J. S. Bell \cite{Bell'1964}. Local outcomes of spatially separated quantum systems prepared in entangled states can produce correlations that can not have any \emph{local realistic} description -- manifesting the nonlocal phenomena. Such nonlocal behavior can be witnessed by violation of some local realistic inequality known as Bell type inequality-- Clauser-Horne-Shimony-Holt (CHSH) inequality is one such celebrated example \cite{Clauser'1969}. CHSH inequality considers the simplest $2-2-2$ scenario that involves two spatially separated parties, each performing two local measurements, each measurement having two outcomes. While the local bound of CHSH expression is $2$, the maximum achievable value of this expression in quantum theory (QT) is  $2\sqrt{2}$, known as Cirel'son bound \cite{Cirel'son'1980}. However, nonlocality is not a salient feature of QT alone. In 1994, Popescu and Rohrlich designed a correlation, famously known as PR correlation, which satisfies the relativistic causality or more broadly no-signaling (NS) principle but at the same time depicts stronger nonlocal behavior as it achieves the algebraic maximum of the CHSH expression \cite{Popescu'1994}. This observation commences a very important question: whether there exists some other fundamental principle(s) (different from no-signaling) limiting the nonlocal strength of QT.

In the last few years several information theoretic as well as physical principles, viz. Non-trivial Communication Complexity \cite{deWolf2010}, Information Causality  \cite{Pawlowski'2009}, Macroscopic Locality  \cite{Navascues2009}, Local Orthogonality \cite{Fritz'2013} have been proposed that successfully explain the limited CHSH violation of QT. These principles also identify a part of the boundary between the set of quantum correlations and the post-quantum NS correlations \cite{Navascues2015,Scarani2009}. Furthermore, applicability of theses principles have also been proved useful in more general $m-n-k$ scenarios. In a different approach, it has been shown that the limited CHSH nonlocality of QT can be connected to other fundamental features of the theory: Heisenberg's uncertainty principle \cite{Oppenheim1072,Zhen2016}, Bohr's complementarity principle \cite{Banik2013,Busch2014} and preparation contextuality \cite{Rai2015}. These connections not only holds true in QT but are also plausible in a larger class of theories.   

A great deal of research has been done on quantum and post-quantum nonlocal correlations in discrete input-output scenario \cite{Pironio2005,Roberts2005,Wolf2014,Nonlocality}. In the quantum domain these studies mainly consider finite-input finite-output correlations arising from finite dimensional quantum systems. Although nonlocal correlations have been studied for infinite dimensional continuous variable (CV) systems \cite{Collett1991,Reid1998,Shchukin,wenger,NhaCarmichael,munro}, those results are fundamentally not different from the finite input-output scenario as discrete binning of continuous outcomes has been considered there. A notable exception, in this regard, is the CV Bell inequalities introduced by Calvalcanti-Foster-Reid-Drummond (CFRD) \cite{Drummond2007}. There the authors derived a class of nonlinear Bell inequalities that apply for continuous outcome spectrum without any need of discrete binning of the outcomes. A natural question of interest in this context will be the notion of post-quantum nonlocal correlations. Very recently, Ketterer et al. have developed a formalism to address this question for generic NS black-box measurement devices with continuous outputs and they have also provided a class of post-quantum nonlocal correlations when only two sites/modes are involved \cite{Aolita2018}. 

A relevant question in the continuous outcome scenario is: how to certify post-quantum nonlocality  of a given correlation? The authors in Ref.\cite{Aolita2018} have used the fact that for two-mode scenario there is no quantum violation of the CFRD inequality \cite{Salles2010}, i.e., CFRD violation at the same time works as nonlocal witness as well as post-quantumness witness. However, this a very specific feature of two-mode case that does not hold for higher number of modes in general \cite{Sallescav}. On the other hand, the principle based methods \cite{Pawlowski'2009,Navascues2009,Fritz'2013} that have been proved to be eminent for studying post-quantum correlations in the discrete outcomes scenario are yet to be generalized for continuous outcome spectrum. Developing a systematic approach to study post-quantum nonlocal correlations for continuous outcome scenario in multi-mode cases is thus quite important. Interestingly we find that Robertson-Schr\"{o}dinger (RS) uncertainty relation has a role to play in this regard. We construct a class of continuous outcome post-quantum nonlocal correlations for generic $m$-mode scenario. While the nonlocality of the proposed class of correlations is certified through violation of CFRD inequalities, post-quantum nature is guaranteed by violation of RS uncertainty relation. 
        
Rest of the paper is organized as follows. In Section II, we briefly revise the framework introduced in \cite{Aolita2018} for CV outcomes. In Sections III and IV, we review CV Bell's inequalities and Robertson-Schr\"{o}dinger uncertainty relation, respectively. Our main results are presented in Section V and finally, we draw our conclusions in Section VI. Some details of the calculations are given in Appendices.

\section{The Framework}
Standard $m-n-k$ Bell scenario considers $m$ space-like separated observers/sites denoted as $\mathcal{A}_i$, with $i\in\{1,...,m\}$. Each observer performs $n$ different measurement $X_i$, with $X_i\in\{0,...,n-1\}$; and each measurement having $k$ distinct outcomes $A_i$, with $A_i\in\{0,...,k-1\}$. Here, likewise in \cite{Aolita2018}, we consider that the outcomes are continuum, i.e., $A_i$ $\in\mathbb R$. While considering continuous outcomes it is convenient to adopt the language of probability measures.
Basic primitive of a probability space consists of three elements: (i) a sample space ($\Omega$), (ii) the Borel $\sigma$-algebra ($\mathcal B(\Omega)$) of events on $\Omega$, and (iii) a valid Borel probability measure $\xi:\mathcal B(\Omega)\rightarrow [0,1]$. In our case, the sample space would be $\Omega=\Omega_1\times\Omega_2\times\cdots\times\Omega_{m}$, with $\Omega_i=\mathbb R$ being the outcome space of $i^{th}$ site. The probability measure satisfies the normalization condition: $\xi(\mathbb{R}\times\mathbb{R}\cdots\times\mathbb{R})=1$, and also satisfies the additivity property: $\xi(\cup_i \omega_i)=\sum_i\xi(\omega_i)$, for all countable sequences $\{\omega_i\}_i$ of disjoint events $\omega_i\in\mathcal{B}(\Omega)$. The relation between a probability measure $\xi$ and a
probability density $p$ is given by,
\begin{eqnarray}\label{den}
&&\xi(A_1\times\cdots\times A_{m}):=\int_{A_1\times \cdots\times A_{m}}d\xi(a_1^\prime,\cdots,a_{m}^\prime)\nonumber\\
&=&\int_{A_1}\cdots\int_{A_{m}} p(a_1^\prime,\cdots,a_{m}^\prime) da_1^\prime\cdots da_{m}^\prime.
\end{eqnarray}
Here $A_1\times\cdots\times A_{m}\in\mathcal{B}(\Omega)$, $A_i\in\mathcal{B}(\mathbb{R})$, $p(a_1^\prime,\cdots,a_{m}^\prime)$ denotes the
corresponding probability density to $\xi$.
We will denote the set of all probability measures on $\mathcal{B}(\Omega)$ as $\mathcal{M}_{\mathbb{R}^m}$.

From now on we consider that one of two possible local measurements will be performed on each site, i.e., $X_i\in\{0,1\},~\forall~i$. In such a scenario, an $m$-mode Bell behavior is defined as the collection of joint
conditional probability measures $\{\xi_{X_1\cdots X_{m}}^{\mathcal{A}_1\cdots\mathcal{A}_m}~|~X_1,\cdots, X_{m}=0,1\}$, where each $\xi_{X_1\cdots X_{m}}^{\mathcal{A}_1\cdots\mathcal{A}_m}\in \mathcal{M}_{\mathbb{R}^m}$. Whenever there is no confusion we will avoid the mode index denoted as superscript. Collection of all $m$-mode Bell behavior will be denoted as $\mathcal{M}_{\mathbb{R}^m}^{2^m}$. Consider any arbitrary grouping of $m$ modes into two disjoint (nonempty) sets $\mathcal{K}$, $\mathcal{K}^{c}$ with $\mathcal{K}\cup\mathcal{K}^{c}= \{\mathcal{A}_1,\cdots,\mathcal{A}_m\}$. NS condition puts the restrictions that measurement choice of one set does not determine the outcome probability of other set for any of the above groupings. In measure theoretic language these conditions read as: 
\begin{align}\nonumber
\xi_{\{X_i\}_{i\in\mathcal{K}}\cup \{X_j\}_{j\in\mathcal{K}^c}}
\left(\prod_{i\in\mathcal{K}}A_i\times\prod_{j\in\mathcal{K}^c}\mathbb{R}_j\right)~~~~~~~~~~~~
\\ = \xi_{\{X_i\}_{i\in\mathcal{K}}\cup \{X_j\oplus 1\}_{j\in\mathcal{K}^c}}
\left(\prod_{i\in\mathcal{K}}A_i\times\prod_{j\in\mathcal{K}^c}\mathbb{R}_j\right), 
\end{align}
for all $A_i\in \mathcal B(\mathbb R)$, where $\oplus$ denotes modulo two sum. The set of all no-signaling correlations $\mathcal{M}_\mathrm{NS}$ is naturally a strict subset of $\mathcal{M}_{\mathbb R^m}^{2^m}$. A behavior will be called quantum \emph{iff} it can be obtained according to Born probability rule, i.e.:
${\xi_{X_1\cdots X_{m}}}(A_1\times\cdots\times A_{m})= \mbox{Tr}\left[\otimes_{i=1}^{m}M_{X_i}{(A_i)}\rho_{\mathcal{A}_1\cdots\mathcal{A}_m} \right],~\forall~A_i\in \mathcal{B}(\mathbb{R})$; where $\rho_{\mathcal{A}_1\cdots\mathcal{A}_m}$ is a density operator acting on some tensor product Hilbert space $\otimes_{i=1}^m\mathcal{H}_i$, with  $\mathcal{H}_i$ being the $i^{th}$ site's Hilbert space (in this case infinite dimensional); and $M_{X_i}{(A_i)}:~\mathcal{B}(\mathbb{R})\mapsto \mathcal{L}_{+}(\mathcal{H}_i)$ are the positive operator valued measures on $\mathcal{H}_i$. A behavior $\{\xi_{X_1\cdots X_{m}}\}_{X_i=0,1}$ will be called post-quantum if $\{\xi_{X_1\cdots X_{m}}\}_{X_i=0,1}\in\mathcal{M}_{NS}$ but $\{\xi_{X_1\cdots X_{m}}\}_{X_i=0,1}\notin\mathcal{M}_{Q}$, the set of quantum behaviors. Local-realistic correlations are those where the outputs are locally generated from local inputs and some pre-established classical correlations encoded in some shared variable $\lambda\in\Lambda$. Such behaviors are of the form $\xi_{X_1\cdots X_{m}}=\int_\Lambda \delta_{a_1(X_1,\lambda),\cdots,a_{m}(X_{m},\lambda)}\, d\eta(\lambda)$, 
where $\eta:\mathcal B(\Lambda)\rightarrow \mathbb R_{\geq0}$ is a probability measure and $\delta_{a_1(x_1,\lambda),\cdots,a_{m}(X_{m},\lambda)}$ is the CV version of the $\lambda$-th local deterministic response function: $\delta_{a_1,\cdots,a_{m}}(A_1\times\cdots\times A_{m}):=1$ if $a_i\in A_i$ and $0$, otherwise. Set of all local behaviors $\mathcal{M}_L$ is a strict subset of $\mathcal{M}_Q$ and  behaviors not belonging to $\mathcal{M}_L$ manifest nonlocal feature.

\section{Continuous Variable Bell's inequalities}
Initial study of Bell test for CV systems was based upon coarse graining of the continuous outcome spectrum into discrete domains \cite{wenger,NhaCarmichael,munro,Acin}.  
One of the main motivations of studying CV Bell scenario is to achieve better detection efficiency as the Homodyne detection method is a highly efficient detection technique \cite{Acin,Munro1999,Cerf2005}. Another way to increase the detection efficiency is to use the idea of continuous realizations of outcomes instead of discrete ones. The idea was initially motivated by the CV version of EPR paradox \cite{Ou1992}. In Ref.\cite{Drummond2007} Cavalcanti, Foster, Reid and Drummond (CFRD) derived a class of local realistic inequalities without any assumption on the number of measurement outcomes and therefore their inequalities are directly applicable to CV systems with no need of discrete
binning of the outcomes. In a multi-site setting each equipped with multiple causally separated apparatus, they considered any real, complex, or vector function $F(\textbf{X}^1,\textbf{X}^2,\cdots)$ of local observables.
All such functions, in a local hidden variable (LHV) theory, are functions of hidden variables $\lambda\in\Lambda$. The average over the LHV ensemble $P(\lambda)$ is given by, $\langle F\rangle=\int_{\Lambda} P(\lambda)F(\textbf{X}^1,\textbf{X}^2,\cdots)d\lambda$. Using the fact that any function of random variables has non-negative variance, the class of CRFD local realistic inequalities read as: $|\langle F\rangle|^2\le \left\langle |F|^2\right\rangle$. For the two-site scenario it was first shown that it is impossible to violate the CFRD inequality with quantum phase-space quadrature operators \cite{Salles2010}. Consequently this result has been generalized for arbitrary quantum measurements \cite{Shchukin}. However it is possible to obtain violation of CFRD inequalities in QT with higher number of modes, in particular, explicit violation has been shown for multipartite GHZ like states \cite{Sallescav}. We will use this particular class of inequalities to establish nonlocal feature of a continuous outcome correlation. Before arriving at our result, let us digress to Robertson-Schr\"{o}dinger (RS) uncertainty relation a bit which plays a crucial role for our purpose.
\begin{figure}[b]
	\begin{center}
		\includegraphics[width=8cm,height=5cm]{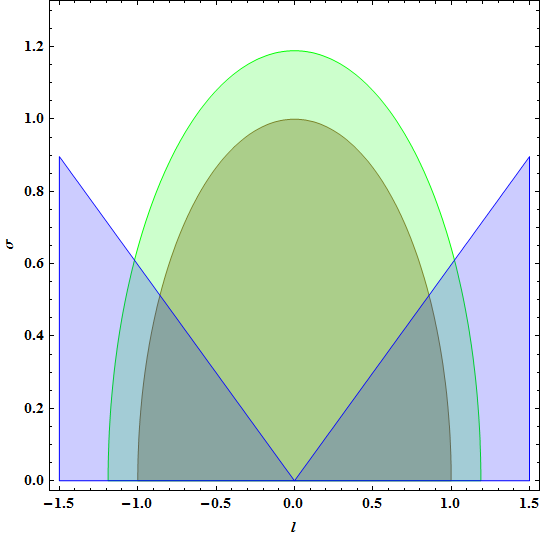}
	\end{center}
	\caption{(Color online)  The region bounded by blue curve denotes the ranges of $(l,\sigma)$ for which $3$-mode CFRD inequality \eqref{result2} is violated. The region bounded by inner half-circle denotes the ranges of $(l,\sigma)$ of \eqref{3m} which violates RS uncertainty relation with the product choice of distribution ($c=0$). The region bounded by green half-circle represents RS violation for non-product choice of distribution (here we have considered $c=1$). The overlapping region of blue curve \& inner half-circle and blue curve \& green curves indicate $3$-mode  post-quantum nonlocal correlations, respectively. Clearly, $c=0$ corresponds to the minimum region violating the RS relation. }
	\label{fig:c3m}
\end{figure}  

\section{Robertson-Schr\"{o}dinger Uncertainty Relation}
A proper mathematical formulation of Heisenberg's preparation uncertainty relation was first introduced by Kennard \cite{Kennard1927}. Schr\"odinger re-derived this idea for two observables correlations in a more refined way \cite{Schrodinger1930} which was further extended for more than two observables by Robertson \cite{Robertson1929}. For $m$-mode quantum state $\rho_{\mathcal{A}_1\cdots\mathcal{A}_m}$, the non-commutativity of the canonical operators and the positive semi-definiteness of the state leads to the famous restriction -- the RS uncertainty relation: $\mathbf{V}+\iota\boldsymbol{\Omega}\geq 0$ \cite{Dutta1994}, 
where $\mathbf{V}$ is a $2m\times 2m$ real symmetric matrix, namely, the covariance matrix (CM) and  
$\boldsymbol{\Omega}$ is the {\it symplectic form} and $\iota=\sqrt{-1}$. CM is calculated from the second moments of position $(\hat{q}_i)$ and momentum $(\hat{p}_i)$ operators which we denote as elements of a vector $\hat{\alpha} = \left(\hat{q}_1,\hat{p}_1,\cdots \hat{q}_{m},\hat{p}_{m}\right)^\intercal$. Then we have, $V_{ij}:=\tfrac{1}{2}\left\langle \left\{  \Delta\hat{\alpha}_{i},\Delta\hat{\alpha}%
_{j}\right\}  \right\rangle_\rho$, where $\Delta\hat{\alpha}_{i}:=\hat{\alpha}_{i}-\langle\hat{\alpha}_{i}\rangle$, $\{.,.\}$ denotes anti-commutator and $\boldsymbol{\Omega}$ is defined as $2\iota\Omega_{ij}=[\hat{\alpha}_i,\hat{\alpha}_j]$. Whether any given real symmetric matrix corresponds to a \textit{bona fide} quantum CM can be verified by RS uncertainty relation. This criterion is necessary and sufficient for Gaussian states while for more general non-Gaussian states it is only a necessary criterion.

\section{Robertson Schr\"{o}dinger uncertainty relation as witness of post-quantumness}
Equipped with all the required tools, we now introduce continuous outcome post-quantum nonlocal correlations for $m$-mode scenario. First we give an example in 3-mode case. Consider the following Bell behavior: 
\begin{subequations}\label{3m}
	\begin{eqnarray}\nonumber
	\xi_{111}^{\mathcal{A}_1\mathcal{A}_2\mathcal{A}_3}&=& \frac{1}{4}\left[\mathcal{N}_{(l,l,-l),\sigma} + \mathcal{N}_{(l,-l,l),\sigma}~~~~~~~~~~~~~~~~\right.\\ &&~~~~~~~~~+\left.\mathcal{N}_{(-l,l,l),\sigma}+ \mathcal{N}_{(-l,-l,-l),\sigma})\right],\\\nonumber
	\xi_{\text{rest}}^{\mathcal{A}_1\mathcal{A}_2\mathcal{A}_3}&=& \frac{1}{4}\left[\mathcal{N}_{(l,l,l),\sigma} + \mathcal{N}_{(l,-l,-l),\sigma}~~~~~~~~~~~~~~~~\right.\\ &&~~~~~~~~~+\left.\mathcal{N}_{(-l,l,-l),\sigma}+ \mathcal{N}_{(-l,-l,l),\sigma}\right],
	\end{eqnarray}
\end{subequations}
where, $\text{rest}\in\{0,1\}^3\setminus\{111\}$, with $0$ and $1$ denoting position and momentum measurements respectively. $\mathcal{N}_{\mathbf{a},\sigma}$ is the normal (Gaussian) probability measure defined through \eqref{den} with probability density centered around $\mathbf{a}:=(a_1,a_2,a_3)$ with width $\sigma$, \textit{i.e.}, $p_{\mathbf{a},\sigma}(\mathbf{a}')= 1/(\sigma\sqrt{2\pi})^3 \exp\left[-\sum_{i=1}^3(a_i-a_i')^2/(2\sigma^2)\right]$. It is straightforward to show that the above behavior is indeed a NS behavior.
CFRD inequality for three modes is defined as \cite{Drummond2007},
\begin{equation}\label{3cfrd}
\left\langle\tilde{X}_{3}\right\rangle^{2}+\left\langle\tilde{Y_{3}}\right\rangle^{2}\leq\left\langle\prod_{k=1}^{3}\left(\left(X_0^{k}\right)^{2}+\left(X_1^{k}\right)^{2}\right)\right\rangle
\end{equation}
where, $\tilde{X}_{3}$ and $\tilde{Y}_{3}$ are obtained from $\tilde{X}_{3}+\iota\tilde{Y}_{3}=\prod_{k=1}^{3}(X_0^{k}+\iota X_1^{k})$. For the correlation \eqref{3m}, the CFRD expression \eqref{3cfrd} turns out to be,
\begin{equation}\label{result2}
	5l^6\leq2\left(l^2+\sigma^2\right)^3.
\end{equation}
For suitable choices of $(l,\sigma)$, correlation (\ref{3m}) can violate inequality-(\ref{result2}), as shown in Fig.\ref{fig:c3m}, and hence establishes nonlocality of those correlations. Naturally the question arises whether such nonlocal correlations are quantum realizable or whether they are post-quantum in nature. One way is to find the 2-mode marginal correlations and check whether the 2-mode marginals violate the 2-mode CFRD inequality. But in this case, the 2-mode marginals being a local correlations satisfy the corresponding CFRD inequality \cite{Appendix}. 

\begin{figure}[b]
	\begin{center}
		\includegraphics[width=8cm,height=5cm]{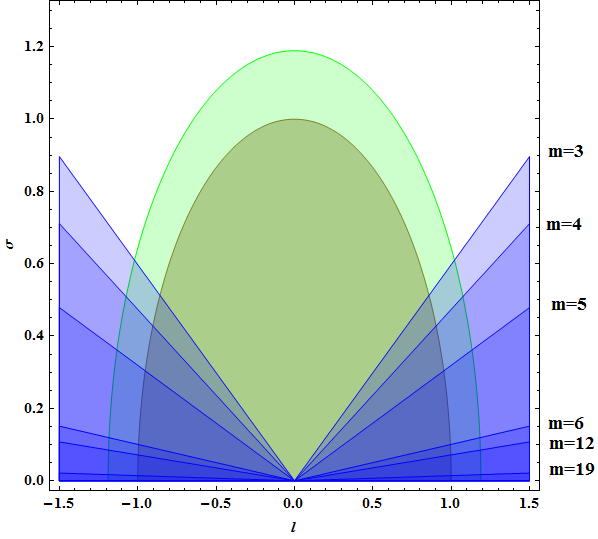}
	\end{center}
	\caption{(Color online) For different number of modes ($m=3,4,5,6,12,19$) the corresponding CFRD inequalities violation has been depicted by different shades of Blue regions. The half circular regions denote RS uncertainty violation as in Fig.\ref{fig:c3m}. For any arbitrary $m$ we do not get violation of CFRD inequalities (eg. $m=7,8,9$). }
	\label{fig:cmm}
\end{figure}

So, at this point we utilize the RS uncertainty relation which puts necessary conditions on a distribution to be quantum realizable: if the RS uncertainty relation is violated then the given distribution can not be a quantum realizable one. To calculate the CM from (\ref{3m}), we require and can readily get the single-mode marginals $\xi^{\mathcal{A}_i}_{X_i}$ as well as the 2-mode marginals $\xi^{\mathcal{A}_i\mathcal{A}_j}_{X_iX_j}$ by integrating out the other mode(s). But calculation of CM also requires single-mode position-momentum joint distribution $\xi^{\mathcal{A}_i}_{(X^i=0,X^i=1)}$. Note that, given marginal probability distributions $\xi^{\mathcal{A}_i}_{X^i=0}$ and $\xi^{\mathcal{A}_i}_{X^i=1}$ the choice of joint distribution $\xi^{\mathcal{A}_i}_{(X^i=0,X^i=1)}$ is not unique. With (trivial) product choice of distribution $\xi^{\mathcal{A}_i}_{(X^i=0,X^i=1)}=\xi^{\mathcal{A}_i}_{X^i=0}\times\xi^{\mathcal{A}_i}_{X^i=1}$ we have $\langle \hat{q}^i\hat{p}^i\rangle=0$, which in turn gives that the RS uncertainty relation will be violated if $l^2+\sigma^2<1$ \cite{Appendix}, \emph{i.e.}, describing a half-circle region on $l-\sigma$ plane [see Fig.\ref{fig:c3m}]. In Fig.\ref{fig:c3m} the overlapping region of blue curve and the inner half-circle violates both the CFRD inequality and the RS relation (calculated with product choice of distribution) and hence establishes post-quantum nonlocality of those correlations. At this point one can ask the question whether the values of $(l,\sigma)$ lying outside the inner circle but withing the blue region denote quantum realizable probability distribution. However it is not straight forward to answer this question. First of all, if we calculate CM with some non product distribution  $\xi^{\mathcal{A}_i}_{(X^i=0,X^i=1)}\neq\xi^{\mathcal{A}_i}_{X^i=0}\times\xi^{\mathcal{A}_i}_{X^i=1}$ we have $\langle \hat{q}^i\hat{p}^i\rangle=c$, with  $c$ being a real number ($c=0$ corresponding to the product choice), and consequently RS uncertainty relation will be violated if $l^2+\sigma^2<\sqrt{1+c^2}$. Therefore the area of post-quantum region increases, as shown in Fig.\ref{fig:c3m} by green half-circle (for $c=1$). Even if one can specify the value of $c$, it will not be possible in general to guarantee quantumness of the correlations outside the green half-circle region as RS relation is a sufficient criterion for bona-fide CM only for Gaussian distribution. However, this calculation asserts the existence of post-quantum nonlocal correlations independent of the fact whether we take product or non-product form of joint position-momentum distribution for single modes. 

We now generalize the above 3-mode example to $m$ number of modes. Consider a vector $\mathbf{P}_i\in\mathbb{R}^m$ with first $i$ number of elements being $-l$ and following $(m-i)$ number of elements being $+l$. Denote by $\mathcal{P}_i$ the set of all vectors obtained from $\mathbf{P}_i$ by permuting its elements. Consider now, an $m$-mode Bell behavior defined as,
\begin{subequations}\label{mm}
	\begin{eqnarray}
	\xi_{11\cdots1}^{\mathcal{A}_0\mathcal{A}_1\cdots\mathcal{A}_m} &= \frac{1}{2^{m-1}}\sum\limits_{\substack{i\in \mathbb{N}_{\text{o}}\\i\leq m}}\sum\limits_{\mathbf{P}_i\in\mathcal{P}_i}\mathcal{N}_{\mathbf{P}_i,\sigma}~,\\
	\xi_{\text{rest}}^{\mathcal{A}_0\mathcal{A}_1\cdots\mathcal{A}_m} &= \frac{1}{2^{m-1}}\sum\limits_{\substack{i\in \mathbb{N}_{\text{e}}\\i\leq m}}\sum\limits_{\mathbf{P}_i\in\mathcal{P}_i}\mathcal{N}_{\mathbf{P}_i,\sigma}~.\\\nonumber
	\end{eqnarray}
\end{subequations}
Here, $\mathbb{N}_o~(\mathbb{N}_e)$ denotes the set of odd (even) integers, and $\mathcal{N}_{\mathbf{a},\sigma}$ is the normal (Gaussian) probability measure defined through \eqref{den} with probability density centered around $\mathbf{a}\equiv(a_1,\cdots,a_m)$ with widths $\sigma$, \textit{i.e.}, $p_{\mathbf{a},\sigma}(\mathbf{a'})= 1/(\sigma\sqrt{2\pi})^m \exp\left[-(\sum_{i=1}^m(a_i-a_i')^2)/(2\sigma^2)\right]$. The expression of $m$-mode CFRD inequality with this probability measure takes the following form \cite{Appendix}: when $m$ is even we get,
\begin{align}\label{nmresullt1a}\nonumber
\left[\left(2^{m/2}\cos(\tfrac{m\pi}{4})+(-1)^{\frac{m}{2}+1}2\right)^2 +2^{m}\sin^2(\tfrac{m\pi}{4})\right]l^{2m} \\
\leq 2^m \left(l^2+\sigma^2\right)^m, ~~~~~~~
\end{align}
For odd $m$, it is found to be,
\begin{align}\label{nmresullt1b}\nonumber
\left[\left(2^{m/2}\sin(\tfrac{m\pi}{4})+(-1)^{\frac{m-1}{2}+1}2\right)^2 +2^{m}\cos^2(\tfrac{m\pi}{4})\right]l^{2m} \\
\leq 2^m \left(l^2+\sigma^2\right)^m. ~~~~~~~~~~~~~~~~~~
\end{align}
For suitable choices of $(l,\sigma)$ $m$-mode probability measure of Eq.(\ref{mm}) violate the corresponding CFRD inequality [see Fig.\ref{fig:cmm}]. And a similar calculation as in the 3-mode example, shows that the RS uncertainly relation, calculated with single mode product [non-product] joint distribution, will be violated by the probability measure Eq.(\ref{mm}) if $l^2+\sigma^2<1~[l^2+\sigma^2<\sqrt{1+c^2}]$. Correspondingly the choices of $(l,\sigma)$ that violates both the CFRD inequality and RS uncertainty relation gives the $m$-mode post-quantum nonlocal correlations.   

So far, we have shown that RS uncertainty relation plays a crucial role in certifying post-quantumness for $m$-mode CV correlations, with $m\ge3$. What will be the implication of our approach for 2-mode case? We find that \cite{Appendix} for the 2-mode case, the probability measure (\ref{mm}), originally considered in \cite{Aolita2018}, yields the CFRD expression as $2l^4-(l^2+\sigma^2)\le 0$. In this case the RS uncertainty relation, calculated with product and non-product single mode joint distribution, will be satisfied if $(l^2+\sigma^2)\ge\sqrt{1+2l^4}$ and $(l^2+\sigma^2)\ge\sqrt{1+l^4+\left(l^2+c^2\right)^2}$ respectively. From these expression it is evident that any such correlation violating CFRD inequality indeed violates RS uncertainty relation. Therefore the post-quantumness of those correlations can be asserted from RS uncertainty relation even without referring to the results of \cite{Salles2010}.

\section{Conclusions}
Usefulness of Robertson-Schr\"{o}dinger uncertainty relation in detecting multi-mode entanglement has already been demonstrated in \cite{Zubairy2009}. On the other hand, the work by Oppenheim and Wehner \cite{Oppenheim1072} is also quite worthy to mention in the context of the present work. In the $2-2-2$ scenario, they have shown that quantum mechanics cannot be more nonlocal with measurements that respect the uncertainty principle in {\it fine-grained} form. To the best of our knowledge, in the continuous outcome scenario the role of uncertainty principle to certify post-quantumness has been explored for the very first time in the present study. Our work also instigates several interesting questions. Will it be the case that any post-quantum nonlocal correlation violates some form(s) of uncertainty principle? Another curious alley to be explored is to construct {\emph genuine} nonlocal inequality for $m$-mode  scenario with continuous outcome, which can be used to certify the inbuilt genuineness of the correlation presented here. Although our work employs uncertainty relation and CFRD inequality to detect post-quantum nonlocal correlation in the continuous outcome scenario, it is demanding to see whether one could do the same with some operational task.

\section{Acknowledgement}
AR would like to acknowledge his visit to The Institute of Mathematical Sciences, Chennai. MB acknowledges support through an INSPIRE-faculty position at S. N. Bose National Centre for Basic Sciences, by the Department of Science and Technology, Government of India. MB would also like to acknowledge the helpful discussion with Ludovico Lami.

\bibliography{Manuscript}

\newpage
\appendix
\begin{widetext}
\section{$3$-mode scenario}

{\bf Calculation of CFRD inequality:} Given a probability measure $\xi$ with probability density $p$ the expectation $\langle \prod_{k}(X_{i_k}^k)^{n_k}\rangle$ can be calculated according to the following rule:
\begin{equation}
\left\langle \prod_{k}\left(X_{i_k}^k\right)^{n_k}\right\rangle:=\int\prod_{k}\left(A_{i_k}^k\right)^{n_k}d\xi.
\end{equation} 
For 3-mode case given probability measure (\ref{3m}) yields:
\begin{eqnarray}
\left\langle \left(X_{i_1}^1\right)^2\left(X_{i_2}^2\right)^2\left(X_{i_3}^3\right)^2\right\rangle &=& \left(l^2+\sigma^2\right)^3, \quad~\forall i_1,i_2,i_3=\{0,1\};\\
\left\langle X_{i_1}^1 X_{i_2}^2 X_{i_3}^3\right\rangle&=& l^3,\quad \mbox{when}~ i_1i_2i_3=0;\\
\left\langle X_{1}^1 X_{1}^2 X_{1}^3\right\rangle &=& -l^3.
\end{eqnarray}
Finally using the above expressions, the CFRD inequality is calculated as \eqref{result2}.\\\\

{\bf Calculation of RS uncertainty relation:}
Denoting position and momentum observable for $i^{th}$ mode as  $(\hat{q}^i,\hat{p}^i)$, the vector $\vec{\alpha}$ for three modes looks:
\begin{equation}
\hat{\alpha} = \left(\hat{q}^1,\hat{p}^1,\hat{q}^2,\hat{p}^2,\hat{q}^3,\hat{p}^3\right)^T \equiv \hat{\alpha}_i\mid_{i=1,\cdots,6}
\end{equation}
The covariance matrix (CM) $V$ is defined as $
V_{ij}=\frac{1}{2}\langle\{ \Delta\hat{\alpha}_i,\Delta\hat{\alpha}_j\}\rangle$, 
where, $\Delta\hat{\alpha}_i = \hat{\alpha}_i-\langle\hat{\alpha}_i\rangle$ and $\{.,.\}$ is anti-commutator. We can find the single-mode and two-mode marginals from \eqref{3m} by integrating out the appropriate modes and they turn out to be,
\begin{eqnarray}\label{3mmarg}
\xi^{\mathcal{A}_i}_{X_i} &=& \tfrac{1}{2}\left[\mathcal{N}_{l,\sigma}+\mathcal{N}_{-l,\sigma}\right], \quad \forall i=\{1,2,3\},~X_i=\{0,1\},\\
\xi^{\mathcal{A}_i,\mathcal{A}_j}_{X_i,X_j} &=& \tfrac{1}{4}\left[\mathcal{N}_{(l,l),\sigma}+\mathcal{N}_{(l,-l),\sigma}+\mathcal{N}_{(-l,l),\sigma}+\mathcal{N}_{(-l,-l),\sigma}\right], \quad \forall\quad i,j=\{1,2,3\};\quad i\neq j;\quad~X_i,X_j=\{0,1\}.
\end{eqnarray}
While calculating terms like $\langle \hat{q}^i\hat{p}^i\rangle$ we require a position-momentum joint probability distribution $\xi^{\mathcal{A}_i}_{(X_i=0,X_i=1)}$ for $i^{th}$ mode. But given $\xi^{\mathcal{A}_i}_{X_i=0}$ and $\xi^{\mathcal{A}_i}_{X_i=1}$ the choice of $\xi^{\mathcal{A}_i}_{(X_i=0,X_i=1)}$ is not unique. First, we consider (trivial) product choice $\xi^{\mathcal{A}_i}_{(X_i=0,X_i=1)}=\xi^{\mathcal{A}_i}_{X_i=0}\times\xi^{\mathcal{A}_i}_{X_i=1}$. In this case $\langle \hat{q}^i\hat{p}^i\rangle=0$ and the CM becomes,
\begin{equation}
\mathbf{V}^{\mathbf{p}}= \bigoplus_{i=1}^3\begin{bmatrix}
l^2+\sigma^2 & 0 \\
0 & l^2+\sigma^2\\
\end{bmatrix}.
\end{equation}
For a non-product choice $\langle \hat{q}^i\hat{p}^i\rangle=c$, where $c$ is some real number. In this case CM becomes, 
\begin{equation}
\mathbf{V}^{\mathbf{np}}= \bigoplus_{i=1}^3\begin{bmatrix}
l^2+\sigma^2 & c \\
c & l^2+\sigma^2\\
\end{bmatrix}.
\end{equation}
A bona-fide CM needs to satisfy RS uncertainty relation $\mathbf{V}+i\boldsymbol{\Omega}\geq0$, where $\Omega=\bigoplus\limits_{i=1}^3 \begin{bmatrix}0 & 1\\-1 & 0\end{bmatrix}$. Respectively, for product and non-product choices the RS uncertainty relation will be violated if
\begin{eqnarray}
l^2+\sigma^2 &<& 1,\quad\quad\quad\quad\quad(\mbox{for product})\label{3-prod}\\
l^2+\sigma^2 &<& \sqrt{1+c^2}\quad(\mbox{for non-product}).\label{3-nprod}
\end{eqnarray}
By comparing Eqs.(\ref{3-prod})-(\ref{3-nprod}) it is obvious that the region of $(l,\sigma)$ violating RS uncertainty relation for product choice is strictly inscribed by the region of $(l,\sigma)$ violating RS uncertainty relation for non-product choice.

\section{$m$-mode scenario}
{\bf Calculation of CFRD inequality:}
For $m$-mode CFRD inequality was defined in terms of the local observables $\{X_0^{k}, X_1^{k}\}$ as, 
\begin{equation}\label{nmcfrd}
\left\langle\tilde{X}_{m}\right\rangle^{2}+\left\langle\tilde{Y_{m}}\right\rangle^{2}\leq\left\langle\prod_{k=1}^{m}\left(\left(X_0^{k}\right)^{2}+\left(X_1^{k}\right)^{2}\right)\right\rangle,
\end{equation}
where, $\tilde{X}_{m}$ and $\tilde{Y}_{m}$ can be obtained from:
\begin{equation}\label{cn}
\tilde{X}_{m}+\iota\tilde{Y}_{m}=\prod_{k=1}^{m}\left(X_0^{k}+\iota X_1^{k}\right).
\end{equation}
The key point while calculating the $m$-mode CFRD inequality for correlation (\ref{mm}) is,
\begin{eqnarray}
\left\langle X_{i_1}^1 X_{i_2}^2 \cdots X_{i_m}^m\right\rangle&=& l^m,\quad\mbox{if}\quad i_1i_2\cdots i_m=0;\\
\left\langle X_{1}^1X_{1}^2\cdots X_{1}^m\right\rangle &=& -l^m;\\
\left\langle \left(X_{i_1}^1\right)^2\left(X_{i_2}^2\right)^2\cdots\left(X_{i_m}^m\right)^2\right\rangle &=& \left(l^2+\sigma^2\right)^m,\quad\forall\quad i_1,\cdots,i_m\in\{0,1\}.
\end{eqnarray}
Thus, the RHS of \eqref{nmcfrd} is readily seen as,
\begin{equation}\label{nmcfrd3}
\left\langle\prod_{k=1}^{m}\left(\left(X_0^{k}\right)^{2}+\left(X_1^{k}\right)^{2}\right)\right\rangle=2^m\left(l^2+\sigma^2\right)^m.
\end{equation}
Calculation of LHS of \eqref{nmcfrd} requires us to know the number of terms with negative signatures in  $\tilde{X}_{m}$ and  $\tilde{Y}_{m}$ which we define as $a_m$ and $b_m$ respectively. $a_m$ and $b_m$ follow recursion relations which can be specify from the following expressing,
\begin{eqnarray}
\tilde{X}_{m}+\iota \tilde{Y}_{m}&=&\prod_{k=1}^{m}\left(X_0^{k}+\iota X_1^{k}\right),\nonumber\\
&=&\prod_{k=1}^{m-1}\left(X_0^{k}+\iota X_1^{k}\right)\left(X_0^{m}+\iota X_1^{m}\right),\nonumber\\
&=&\left(\tilde{X}_{m-1}+\iota \tilde{Y}_{m-1}\right)\left(X_0^{m}+\iota X_1^{m}\right),\nonumber\\
&=&\left(\tilde{X}_{m-1}X_0^{m}-\tilde{Y}_{m-1}X_1^{m}\right)+\iota \left(\tilde{X}_{m-1}X_1^{m}+\tilde{Y}_{m-1}X_0^{m}\right),\nonumber\\\nonumber\\
\Rightarrow\quad \tilde{X}_{m}&=&\left(\tilde{X}_{m-1}X_0^{m}-\tilde{Y}_{m-1}X_1^{m}\right),\quad\&\quad \tilde{Y}_{m}=\left(\tilde{X}_{m-1}X_1^{m}+\tilde{Y}_{m-1}X_0^{m}\right).
\end{eqnarray}
Thus we have the following coupled recursion relations,
\begin{align}
a_m &=2^{m-2}+a_{m-1}-b_{m-1},\\
b_m &=a_{m-1}+b_{m-1}.
\end{align} 
Closed form expressions for $a_m$ and $b_m$ turns out to be,
\begin{align}
a_m &= \frac{1}{2}\left[2^{m-1}-2^{m/2}\cos(\tfrac{m\pi}{4})\right],\\
b_m &= \frac{1}{2}\left[2^{m-1}-2^{m/2}\sin(\tfrac{m\pi}{4})\right].
\end{align}
We also need to know the signature of the term $X^1_1 X^2_1\cdots X^m_1$ as well as whether it is included in $\tilde{X}_{m}$ or $\tilde{Y}_{m}$. We notice that,
\begin{align*}
(-1)^{m/2} X_1^1X_1^2\cdots X_1^m \in \tilde{X}_m, \text{ if m is even,}\\
(-1)^{(m-1)/2} X_1^1X_1^2\cdots X_1^m \in \tilde{Y}_m, \text{ if m is odd.}
\end{align*}
The required expectation values of $\tilde{X}_{m}$ and  $\tilde{Y}_{m}$ thus become: 
\begin{align}\nonumber\label{nmcfrd1}
\braket{\tilde{X}_{m}}&=\left[2^{m-1}-2a_m+(-1)^{\frac{m}{2}+1}2\right]l^{m},\\
\braket{\tilde{Y}_{m}}&=\left[2^{m-1}-2b_m\right]l^{m},\quad \quad\quad\quad\quad\quad\quad\quad\mbox{for even}\quad m;\nonumber\\\nonumber\\
\braket{\tilde{X}_{m}}&=\left[2^{m-1}-2a_m\right]l^{m},\\
\braket{\tilde{Y}_{m}}&=\left[2^{m-1}-2b_m+(-1)^{\frac{m-1}{2}+1}2\right]l^{m}\quad\quad\quad\mbox{for odd}\quad m.
\end{align}
Finally the CFRD inequality becomes,
\begin{align}\label{anmresullt1a}
\left[\left(2^{m/2}\cos(\tfrac{m\pi}{4})+(-1)^{\frac{m}{2}+1}2\right)^2 +2^{m}\sin^2(\tfrac{m\pi}{4})\right]l^{2m}
\leq 2^m \left(l^2+\sigma^2\right)^m,\quad (\mbox{for even});\\
\left[\left(2^{m/2}\sin(\tfrac{m\pi}{4})+(-1)^{\frac{m-1}{2}+1}2\right)^2 +2^{m}\cos^2(\tfrac{m\pi}{4})\right]l^{2m}\leq 2^m \left(l^2+\sigma^2\right)^m,\quad (\mbox{for odd}). 
\end{align}

{\bf Calculation of RS uncertainty relation:}
As in the 3-mode case, for general $m$-mode case also we have single-mode and two-mode marginals of the following forms,
\begin{eqnarray}\label{3mmmarg}
\xi^{\mathcal{A}_i}_{X_i} &=& \tfrac{1}{2}\left[\mathcal{N}_{l,\sigma}+\mathcal{N}_{-l,\sigma}\right], \quad \forall i=\{1,\cdots,m\},~X_i=\{0,1\},\\
\xi^{\mathcal{A}_i,\mathcal{A}_j}_{X_i,X_j} &=& \tfrac{1}{4}\left[\mathcal{N}_{(l,l),\sigma}+\mathcal{N}_{(l,-l),\sigma}+\mathcal{N}_{(-l,l),\sigma}+\mathcal{N}_{(-l,-l),\sigma}\right], \quad \forall\quad i,j=\{1,\cdots,m\};\quad i\neq j;\quad~X_i,X_j=\{0,1\}.
\end{eqnarray}
As in the 3-mode case, the product and non-product choices of single-mode position-momentum joint distributions give the respective CM matrices:  $\mathbf{V}^{\mathbf{p}}= \bigoplus_{i=1}^m\begin{bmatrix}l^2+\sigma^2 & 0 \\
0 & l^2+\sigma^2\\\end{bmatrix}$ and $\mathbf{V}^{\mathbf{np}}=\bigoplus_{i=1}^m\begin{bmatrix}l^2+\sigma^2 & c \\c & l^2+\sigma^2\\\end{bmatrix}$. And consequently the RS uncertainty will be violated if $l^2+\sigma^2 < 1$ and $l^2+\sigma^2 < \sqrt{1+c^2}$ respectively. 

\section{$2$-mode scenario}
Consider the $2$-mode correlation introduced in Ref.\cite{Aolita2018}:
\begin{subequations}
	\begin{eqnarray*}
		\xi_{00}^{\mathcal{A}_1\mathcal{A}_2}  =\xi_{01}^{\mathcal{A}_0\mathcal{A}_1} =\xi_{10}^{\mathcal{A}_0\mathcal{A}_1}=\frac{1}{2}\left[ \mathcal{N}_{(l,l),\sigma} + \mathcal{N}_{(-l,-l),\sigma}\right],\\
		\xi_{11}^{\mathcal{A}_1\mathcal{A}_2} = \frac{1}{2}\left[ \mathcal{N}_{(l,-l),\sigma} + \mathcal{N}_{(-l,l),\sigma}\right].
	\end{eqnarray*}
\end{subequations}
In this case the CFRD inequality turns out to be: $8l^4\leq 4(l^2+\sigma^2)^2$.
With product and non-product choices of single-mode position-momentum joint distribution the CM becomes,
$$\mathbf{V}^{\mathbf{p}}= \begin{bmatrix}
l^2+\sigma^2 & 0 & l^2 & l^2\\
0 & l^2+\sigma^2 & l^2 & -l^2\\
l^2 & l^2 & l^2+\sigma^2 & 0\\
l^2 & -l^2 & 0 & l^2+\sigma^2
\end{bmatrix};\quad\quad \mathbf{V}^{\mathbf{np}}= \begin{bmatrix}
l^2+\sigma^2 & c & l^2 & l^2\\
c & l^2+\sigma^2 & l^2 & -l^2\\
l^2 & l^2 & l^2+\sigma^2 & c\\
l^2 & -l^2 & c & l^2+\sigma^2
\end{bmatrix}.$$
Respectively, the RS uncertainty relation will be violated if,
\begin{align}
(l^2+\sigma^2)&< \sqrt{(1+2l^4)}\quad\quad\quad\quad\quad\quad(\mbox{for product}),\\
(l^2+\sigma^2)&<\sqrt{1+l^4+(l^2+c^2)^2}\quad(\mbox{for non-product}).
\end{align}
From the expressions of CFRD inequality and RS uncertainty relation it is evident that any $(l,\sigma)$ that violates CFRD inequality also violates RS uncertainty relation (both the product and non-product forms) as shown in Fig.\ref{fig:violation4a}. 

\begin{figure}[h!]
	\begin{center}
		\includegraphics[width=5.5cm,height=5.5cm]{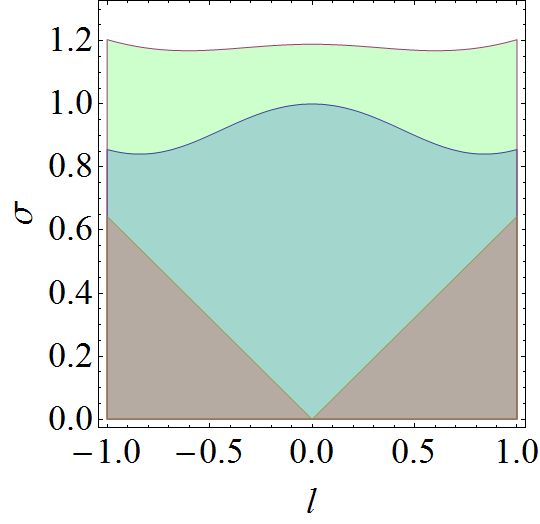}\quad\quad\quad\quad\quad\quad\includegraphics[width=5.5cm,height=5.5cm]{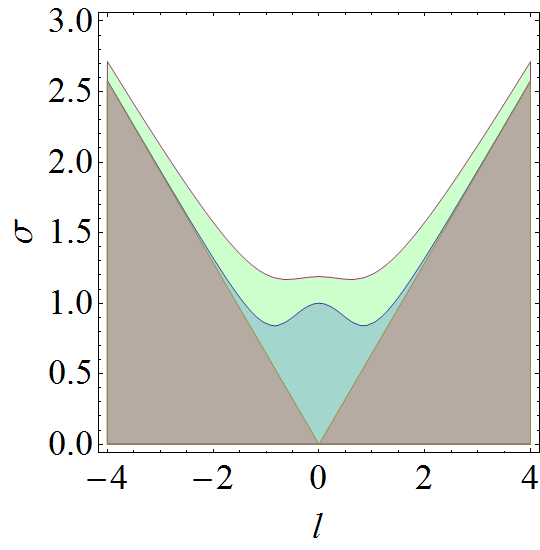}
	\end{center}
	\caption{(Color online) Brown region denotes violation of CFRD inequality. Deep and light green regions correspond to violation of RS uncertainty relation for product and non-product (with $c=1$) choices of single mode position-momentum joint distribution.}
	\label{fig:violation4a}
\end{figure}  

\end{widetext}
\end{document}